\documentclass[twocolumn]{aastex62}

\graphicspath{{./}{figures/}}

\received{\today}
\revised{\today}
\accepted{\today}
\submitjournal{ApJL}

\shorttitle{Tidal spin up and black hole kicks}
\shortauthors{Stevenson}

\begin{document}

\title{Biases in estimates of black hole kicks from the spin distribution of binary black holes}

\author[0000-0002-6100-537X]{Simon Stevenson}
\affil{Centre for Astrophysics and Supercomputing, Swinburne University of Technology, John St, Hawthorn, Victoria- 3122, Australia}
\affil{The ARC Centre of Excellence for Gravitational Wave Discovery,  OzGrav}
\affil{ARC Discovery Early Career Research Award Fellow}



\begin{abstract}

A population of more than 50 binary black hole mergers has now been observed by the LIGO and Virgo gravitational-wave observatories. 
While neutron stars are known to have large velocities associated with impulsive \textit{kicks} imparted to them at birth in supernovae, whether black holes receive similar kicks, and of what magnitude, remains an open question.
Recently, \citet{Callister:2020vyz} analysed the binary black hole population under the hypothesis that they were all formed through isolated binary evolution and claimed that large black hole kicks (greater than 260\,km/s at 99\% confidence) were required for the spin distribution of merging binary black holes to match observations. 
Here we highlight that a key assumption made by \citet{Callister:2020vyz}---that all secondary black holes can be tidally spun up---is not motivated by physical models, and may lead to a bias in their estimate of the magnitudes of black hole kicks. 
We make only minor changes to the \citet{Callister:2020vyz} model, accounting for a population of wider merging binaries where tidal synchronisation is ineffective.
We show that this naturally produces a bimodal spin distribution for secondary black holes, and that the spin-orbit misalignments observed in the binary black hole population can be explained by more typical black hole kicks of order 100\,km/s, consistent with kicks inferred from Galactic X-ray binaries containing black holes.
We conclude that the majority of the binary black hole population is consistent with forming through isolated binary evolution.

\end{abstract}

\keywords{black hole -- gravitational wave -- supernova}


\section{Introduction} 
\label{sec:intro}

A population of more than 50 binary black hole mergers has now been observed by the LIGO \citep{TheLIGOScientificDetector:2014jea} and Virgo \citep{TheVirgoDetector:2014hva} gravitational-wave observatories, as published in a series of catalogues \citep{LIGOScientific:2018mvr,Abbott:2020niy,LIGOScientific:2021usb,GWTC-3}. 

The dominant formation scenario for binary black holes remains uncertain. 
One proposed scenario is the isolated evolution of a binary of two massive stars \citep[][]{Belczynski:2016obo,Stevenson:2017tfq}.
In this channel, two massive stars live their lives together, which interact through tides, phases of mass transfer and common envelope evolution that shrink the orbit of the binary, finally forming a binary black hole that can merge due to the emission of gravitational waves within the age of the Universe. 
Binary black holes formed through isolated binary evolution are expected to have their spins roughly aligned with the binary orbital angular momentum \citep[e.g.,][]{Rodriguez:2016vmx,Stevenson:2017dlk,Gerosa:2018wbw}.
Alternate formation channels such as the dynamical formation of binary black holes in dense stellar environments such as young star clusters or globular clusters have also been proposed \citep{Sigurdsson:1993Natur,Rodriguez:2019huv,DiCarlo:2020lfa}; see \citet{Mandel:2018hfr} for a recent overview of binary black hole formation scenarios.

Several stages of massive binary evolution are poorly constrained. 
For example, we know from radio observations of isolated Galactic pulsars that neutron stars typically have large space velocities of a few hundred km/s \citep{LyneLorrimer:1994Nature,Hobbs:2005yx} that are thought to be imparted at birth in a supernova \citep[][]{Janka:2016nak}.
However, it is still unclear what kicks black holes receive at formation (if any). 
Observations of the kinematics of Galactic X-ray binaries containing black holes have identified that some black holes do receive kicks comparable to neutron stars \citep{Brandt:1995MNRAS,Mirabel:2001Nature}, whilst other systems appear to have much smaller velocities, and some may have received no kick at all \citep[e.g.,][]{Mirabel:2003Sci}.
Fitting the distribution of observed black hole kicks typically suggests that black holes should receive kicks comparable to, or smaller in magnitude than neutron stars \citep{Repetto:2012MNRAS,Repetto:2015kra,Mandel:2015eta,Repetto:2017MNRAS,Atri:2019fbx}.
The reduction may be due to mass falling back onto the proto-compact object during black hole formation \citep[e.g.,][]{Fryer:2012ApJ,Janka:2013hfa,Chan:2017tdg}.
One should also keep in mind that these samples are likely strongly biased by both observational and evolutionary selection effects, and thus they may not represent the true distribution of black hole natal kicks.
The typical magnitude of black hole kicks is an important factor in determining the merger rate of binary black holes \citep[e.g.,][]{Wysocki:2017isg}, as well as the typical angles by which the spins of the black holes are misaligned from the orbital angular momentum \citep[][]{Rodriguez:2016vmx}. 

Recently, \citet{Callister:2020vyz} analysed the distribution of spins of merging binary black holes from \citet{Abbott:2020niy} with a phenomenological model motivated by the isolated binary evolution formation channel and claimed that large kicks (greater than 260\,km/s at 99\% confidence) are required in order to reconcile the predictions of the model with the observations\footnote{\citet{OShaughnessy:2017eks} also inferred that a kick greater than 50\,km/s was required to explain the spin of GW151226 \citep{LIGOScientific:2016sjg}}.
Such large kicks are at odds with other estimates of black hole kick magnitudes, leading \citet{Callister:2020vyz} to suggest that other channels may contribute to the formation of binary black holes, such as dynamical formation in dense star clusters \citep[e.g.,][]{Rodriguez:2019huv,DiCarlo:2020lfa}. 

In this \emph{Letter} we show that making similar assumptions to \citet{Callister:2020vyz} but accounting for the population of wide merging binaries where tidal synchronisation is ineffective yields a bimodal spin magnitude distribution for the second born black hole in merging binary black holes, that agrees well with the observed distribution.
We show that there is currently no evidence for large ($\sim 1000$\, km/s) black hole natal kicks, and that typical kicks of order $\sim 100$\,km/s (consistent with those inferred from Galactic X-ray binaries, as discussed above) are sufficient to produce the observed spin-orbit misalignments.
We therefore argue that (apart from some exceptional events) the majority of the binary black hole population is consistent with forming through isolated binary evolution.

\section{Formation of the second black hole}
\label{sec:second_black_hole}

We begin by constructing a simple model that matches the assumptions made by \citet{Callister:2020vyz} as closely as possible. 
We briefly describe the assumptions of this model in Section~\ref{subsec:setup}.
We then describe in detail the assumptions made in determining the spin of the secondary black hole in Section~\ref{subsec:tidal_spin_up}.

\subsection{Model set up}
\label{subsec:setup}

\citet{Callister:2020vyz} use a phenomenological model
motivated by the classical isolated binary evolution channel. 
In particular, they model the population of black hole-helium star (BH + He star) binaries, the immediate progenitors to binary black holes.
They use this model to perform a Bayesian hierarchical analysis \citep{Mandel:2019MNRAS} of the spin distribution of the binary black hole population from \citet{Abbott:2020niy}, where they fit several free parameters including the typical kicks that black holes receive at formation.
\citet{Callister:2020vyz} assume that all black holes receive a kick at formation, with the black hole kick velocity distribution characterised as a Maxwellian distribution with a dispersion $\sigma$.
In their fitting, they find that values of $\sigma$ as high as $1000$\,km/s are routinely required to induce large enough spin tilts to match the observed spin distribution.

The parameters of the BH + He star binary are the mass of the primary black hole ($m_1$), the mass of the helium star prior to collapse ($M_\mathrm{He}$) and the orbital separation of the binary at the time of the collapse of the He star ($a$).
We follow 
\citet{Callister:2020vyz} in assuming that $m_1$ is drawn from a power-law distribution $p(m_1) \sim m_1^{-2.2}$ between 5\,M$_\odot$ and 75\,M$_\odot$, that $M_\mathrm{He}$ is drawn uniformly between 5\,M$_\odot$ and $m_1$ and that the separation $a$ is drawn from a log-uniform distribution between 5\,R$_\odot$ and 300\,R$_\odot$ based on results from \citet{Bavera:2019}. We assume that these binaries initially have circular orbits.

In order to calculate the mass of the second formed black hole $m_2$, \citet{Callister:2020vyz} assume that helium stars lose around 10\% of their mass in the collapse to a black hole. 
We instead assume that the final black hole mass is equal to the progenitor helium star mass, allowing for no mass loss during the collapse to a black hole, approximating the $\sim 0.1$\,M$_\odot$ expected to be lost through neutrinos \citep[][]{Stevenson:2019rcw,Mandel:2020qwb}. We do not expect this difference in assumption to have a large impact on our results.

Since it is the focus of this letter, we discuss the assumption made regarding the spin of the secondary below.

\subsection{Tidal spin up of secondary}
\label{subsec:tidal_spin_up}

\begin{figure}
    \centering
    \includegraphics[width=\columnwidth]{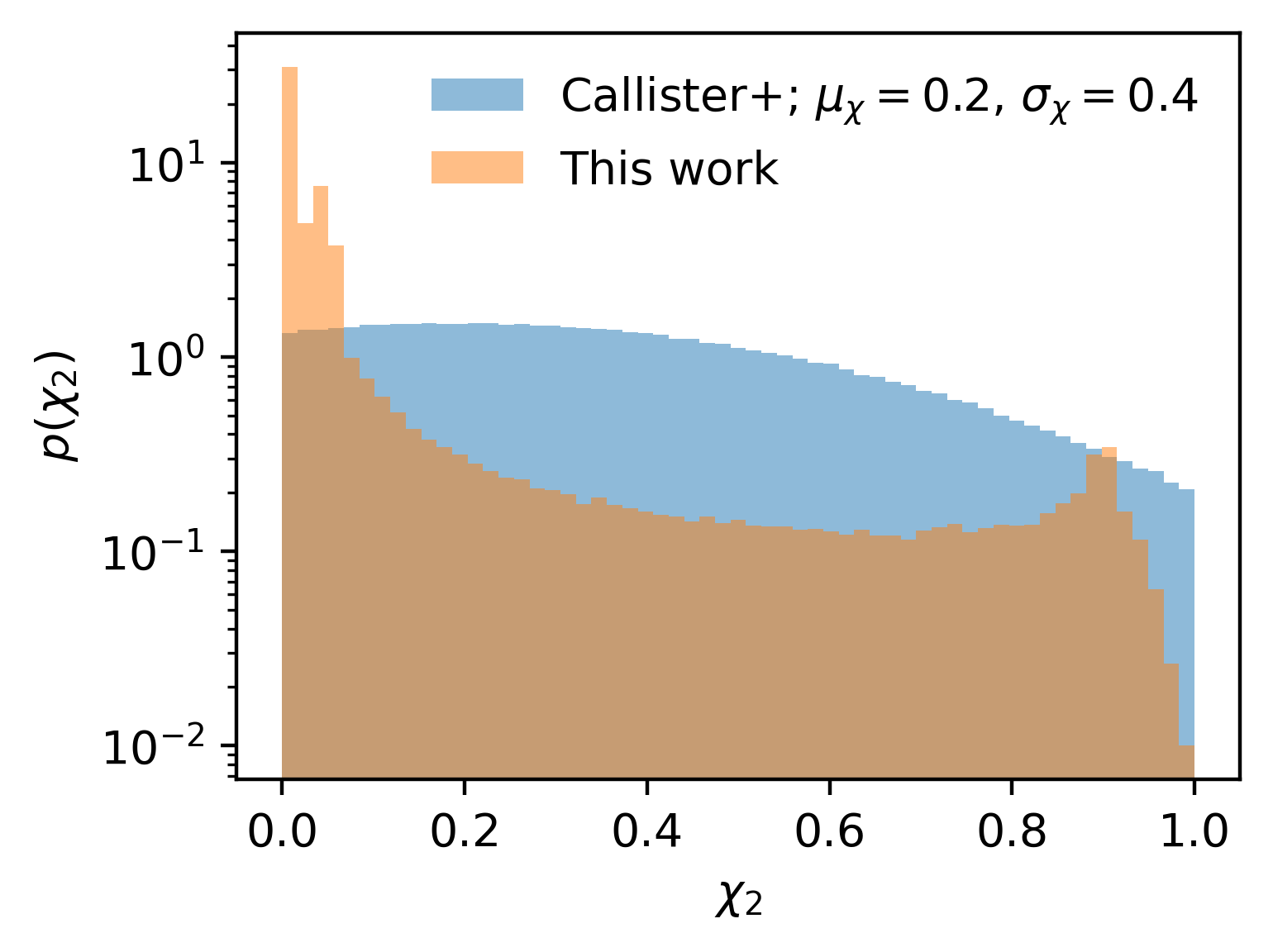}
    \caption{Normalised probability distributions for the secondary spin magnitude $\chi_{2}$. 
    The blue shows a typical result from the analysis of \citet{Callister:2020vyz} when assuming large black hole natal kicks ($\mu_\chi = 0.2$, $\sigma_\chi = 0.4$), whilst in orange we show our model accounting for inefficient tidal synchronisation in wide binaries using the fits from \citet{Bavera:2021evk}. 
    Binary evolution naturally predicts a bimodal secondary spin magnitude distribution which is not well captured by a Gaussian.}
    \label{fig:chi_2}
\end{figure}

Recent models of the binary evolution channel have made important predictions for the spins of the black holes.
In particular, models invoking efficient angular momentum transport in massive stars \citep[e.g.,][]{Spruit:2001tz,Fuller:2019MNRAS} argue that the first black hole is born with a small dimensionless spin magnitude (we assume $\chi_1 = 0$) because the majority of its progenitor's angular momentum was stored in its envelope, and was subsequently removed during the first mass transfer stage \citep{Qin:2018vaa,Fuller:2019sxi}. 
As discussed above, the immediate progenitor of a binary black hole is a tight binary consisting of a helium star and a black hole.
Previous phases of mass transfer, tides and common envelope evolution are expected to have aligned the spins of the binary with the orbital angular momentum prior to this stage \citep[e.g.,][]{Gerosa:2018wbw}.
Tides exerted on the helium star by the black hole can lead to the rotation of the helium star becoming synchronised with the orbital period, potentially leading to a rapidly rotating helium star, that may subsequently collapse to form a rapidly rotating black hole \citep{Detmers:2008A&A,Kushnir:2016zee,Hotokezaka:2017ApJ,2018MNRAS.473.4174Z,Qin:2018vaa,Bavera:2019,Belczynski:2017gds,Steinle:2020xej,Fuller:2022ysb}.
\citet{Callister:2020vyz} assume that this tidal spin up is possible for any orbital separation. 
In their model, the spin magnitude of the second formed black hole $\chi_2$ is drawn from a Gaussian with mean $\mu_\chi$ and standard deviation $\sigma_\chi$ (truncated such that $0 < \chi_2 < 1$); a typical value well supported by the analysis of \citet{Callister:2020vyz} when allowing for large black hole kicks is $\mu_\chi = 0.2$ and $\sigma_\chi = 0.4$. 
We show this distribution in Figure~\ref{fig:chi_2}.

In our analysis we take into account the fact that tidal synchronisation is expected to be ineffective at spinning up helium stars in binaries with orbital periods greater than around 1\,day \citep{Qin:2018vaa,Bavera:2021evk}.
To model this, we use the fitting formulae from \citet{Bavera:2021evk}, which are calibrated to a large set of detailed BH + He star binary models, and provide an estimate of the final spin of the second formed black hole as a function of the binary orbital period and the masses of the helium star and the black hole.
We include only systems that merge due to the emission of gravitational waves in less than the age of the Universe \citep{Peters:1964}, as estimated using the fit from \citet{Mandel:2021RNAAS}.
This selects only binaries with orbital periods less than a few days.

Unlike \citet{Callister:2020vyz}, we start by assuming that black holes receive no kicks at formation, as might be expected for black holes formed through almost complete fallback \citep[e.g.,][]{Chan:2017tdg},
assuming that black hole kicks are generated through the asymmetric ejection of mass during the supernova \citep{Janka:2013hfa}.
As we will show, contrary to the claims of \citet{Callister:2020vyz} that large black hole natal kicks are required to match the observed black hole spin distribution, even this extreme assumption can explain the observed black hole spin distribution. 
Results from this model are discussed below in Section~\ref{sec:spin_dist}.
We discuss implications for black hole kicks in Section~\ref{sec:implications}.

\section{Effective Spin distribution of binary black holes}
\label{sec:spin_dist}

We show the distribution for the spin magnitude of the secondary black hole, $\chi_2$ predicted by our model in Figure~\ref{fig:chi_2}. 
We see that in this model most second born black holes are also slowly rotating with $\chi_{2} \sim 0$, but there is a tail of systems with tight enough orbits that tides can efficiently spin up the helium star, resulting in a second peak around $\chi_{2} \sim 1$ \citep[see also][]{Bavera:2019}.
It is clear by eye that a Gaussian distribution is a poor description of the spin magnitude distribution predicted by the tidal synchronisation model, regardless of the parameters chosen for the Gaussian. 

Comparisons to gravitational-wave observations are often made using the well-measured effective spin parameter \citep[e.g.,][]{Farr:2017uvj,Abbott:2020niy,Callister:2020vyz}
\begin{equation}
    \chi_\mathrm{eff} = \frac{\chi_1 \cos \theta_1 + q \chi_2 \cos \theta_2}{(1 + q)} ,
    \label{eq:chi_eff}
\end{equation}
where $q = m_2/m_1 < 1$ is the binary mass ratio, $\chi_i$ are the dimensionless spin magnitudes of the two black holes and $\cos \theta_i$ are the spin tilt angles between the spins of the black holes and the orbital angular momentum.
Since we assume that black holes form with no associated kick, we also assume that the spin of the secondary is perfectly aligned with the orbital angular momentum, $\cos \theta_2 = \cos \theta_1 = 1$.
Small kicks would result in only minor misalignments \citep{Rodriguez:2016vmx,Stevenson:2017dlk,Gerosa:2018wbw}. 
We discuss the implications for the magnitude of black hole kicks in Section~\ref{sec:implications}.

\begin{figure}
    \centering
    \includegraphics[width=\columnwidth]{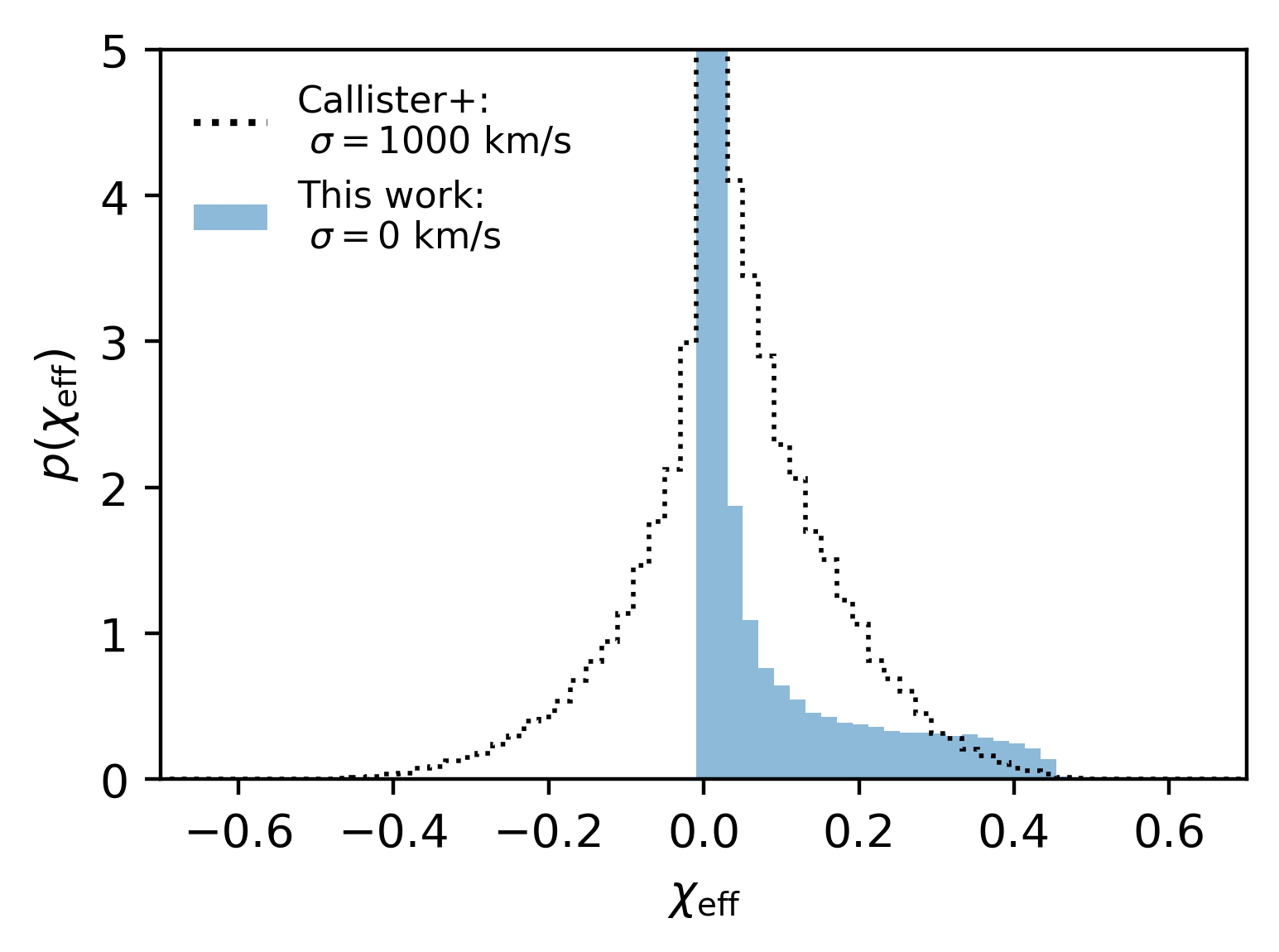}
    \includegraphics[width=\columnwidth]{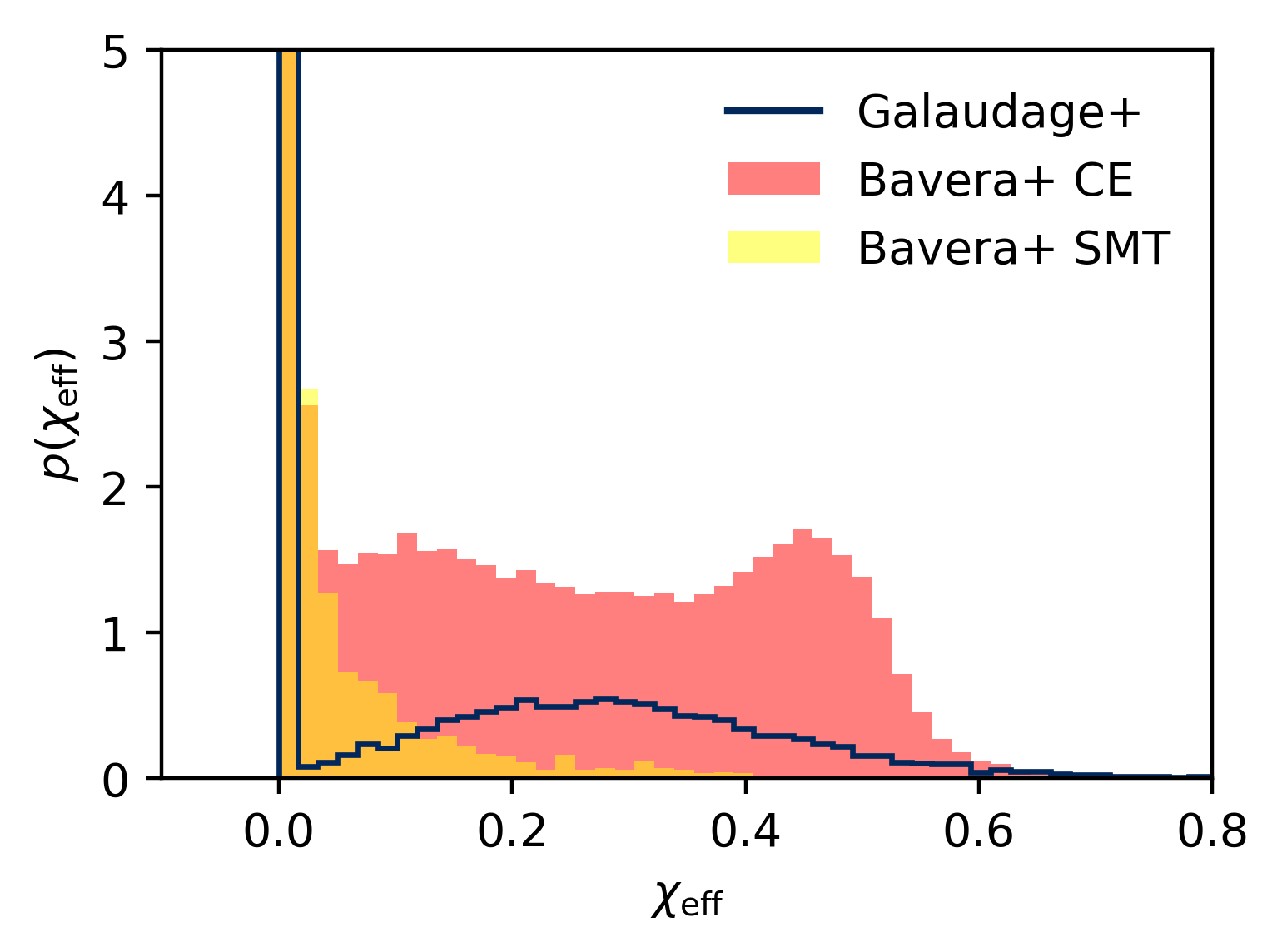}
    \caption{Normalised probability distributions for the effective spin parameter $\chi_\mathrm{eff}$. 
    In the top panel, the blue distribution shows the result from our modified version of the \citet{Callister:2020vyz} model described in Section~\ref{sec:second_black_hole} when assuming no black hole kicks ($\sigma = 0$\,km/s).
    Wide binaries result in a peak at $\chi_\mathrm{eff} = 0$, whilst in tighter binaries the secondary can be tidally spun up, resulting in $\chi_\mathrm{eff} > 0$.
    The dotted black line shows the result when assuming large black hole kicks ($\sigma = 1000$\,km/s) and that the spin magnitude of the second born black hole ($\chi_{2}$) is given by the Gaussian distribution assumed by \citet{Callister:2020vyz} and shown in Figure~\ref{fig:chi_2}.
    This symmetric distribution of $\chi_\mathrm{eff}$ is a typical result found by \citet{Callister:2020vyz}, see their Figure 3.
    The bottom panel shows two example distributions from the population synthesis models from \citet{Bavera:2020uch}; the stable mass transfer channel (SMT; yellow) and the common envelope channel (CE; red), demonstrating how the results of this model may change with different assumptions about the masses and orbital periods of BH + He star binaries.
    We also show the fit to the gravitational-wave observations from \citet{Galaudage:2021rkt} in dark blue. 
    The y-axis is arbitrarily cut-off in both plots for clarity.
    }
    \label{fig:chi_eff}
\end{figure}

We show the distribution of effective spins  $\chi_\mathrm{eff}$ under these assumptions in Figure~\ref{fig:chi_eff}.
The distribution is strongly peaked at $\chi_\mathrm{eff} = 0$ ($\sim 80$\% of the population has $\chi_\mathrm{eff} < 0.05$) with a tail of around 20\% of the population having positive $\chi_\mathrm{eff}$, up to a maximum of around $\chi_\mathrm{eff} = 0.5$, with no support for negative $\chi_\mathrm{eff}$ in agreement with observations \citep{Roulet:2021hcu,Galaudage:2021rkt}.
For comparison, we also show the distribution of $\chi_\mathrm{eff}$ when using assumptions that more closely match the findings of \citet{Callister:2020vyz}; in particular, we assume that the spin of the second born black hole ($\chi_{2}$) is drawn from the Gaussian distribution shown in Figure~\ref{fig:chi_2}, and we assume that black hole kicks are drawn from a Maxwellian distribution with $\sigma = 1000$\,km/s (see Section~\ref{sec:implications} for details). This broadly reproduces the approximately symmetric $\chi_\mathrm{eff}$ distribution found by \citet{Callister:2020vyz}, as shown in their Figure 3.

Whilst we do not explicitly compute the likelihood of the data under the model discussed here, \citet{Galaudage:2021rkt} found a $\chi_\mathrm{eff}$ distribution with a similar shape (shown in the bottom panel of Figure~\ref{fig:chi_eff}) to that found in our modified version of the \citet{Callister:2020vyz} model.
The analysis of \citet{Galaudage:2021rkt} provides a substantially better match (as quantified by the Bayesian evidence) to the gravitational-wave observations than the \textsc{Default} spin model \citep[][]{Talbot:2017PhRvD,Wysocki:2019PRD} used by \citet{LIGOScientific:2020kqk,GWTC-3Astro}. 
We expect that a similar preference would be found compared to the model of \citet{Callister:2020vyz}.
\newpage

\section{Implications for black hole kicks}
\label{sec:implications}

In the previous section we have shown that the phenomenological model for binary black hole formation from \citet{Callister:2020vyz} can be easily modified to more closely match the predictions of detailed binary evolution models (see lower panel of Figure~\ref{fig:chi_eff}).
This modified model produces a distribution for the effective spin parameter $\chi_\mathrm{eff}$ that 
reproduces some of the features of the \citet{Galaudage:2021rkt} fit to current gravitational-wave observations.
Some possible reasons for remaining differences are discussed in Section~\ref{sec:conclusion}.
We therefore argue that a model of binary evolution in which: (a) the progenitor of the second formed black hole in tight binaries are spun up through tides; (b) black holes formed in wide binaries are not spun up and have $\chi_2 \sim 0$, and (c) black holes are formed with no associated kicks, naturally leads to an effective spin distribution that is compatible with gravitational-wave observations, and cannot currently be ruled out based solely on the effective spin distribution. 

\begin{figure}
    \centering
    \includegraphics[width=\columnwidth]{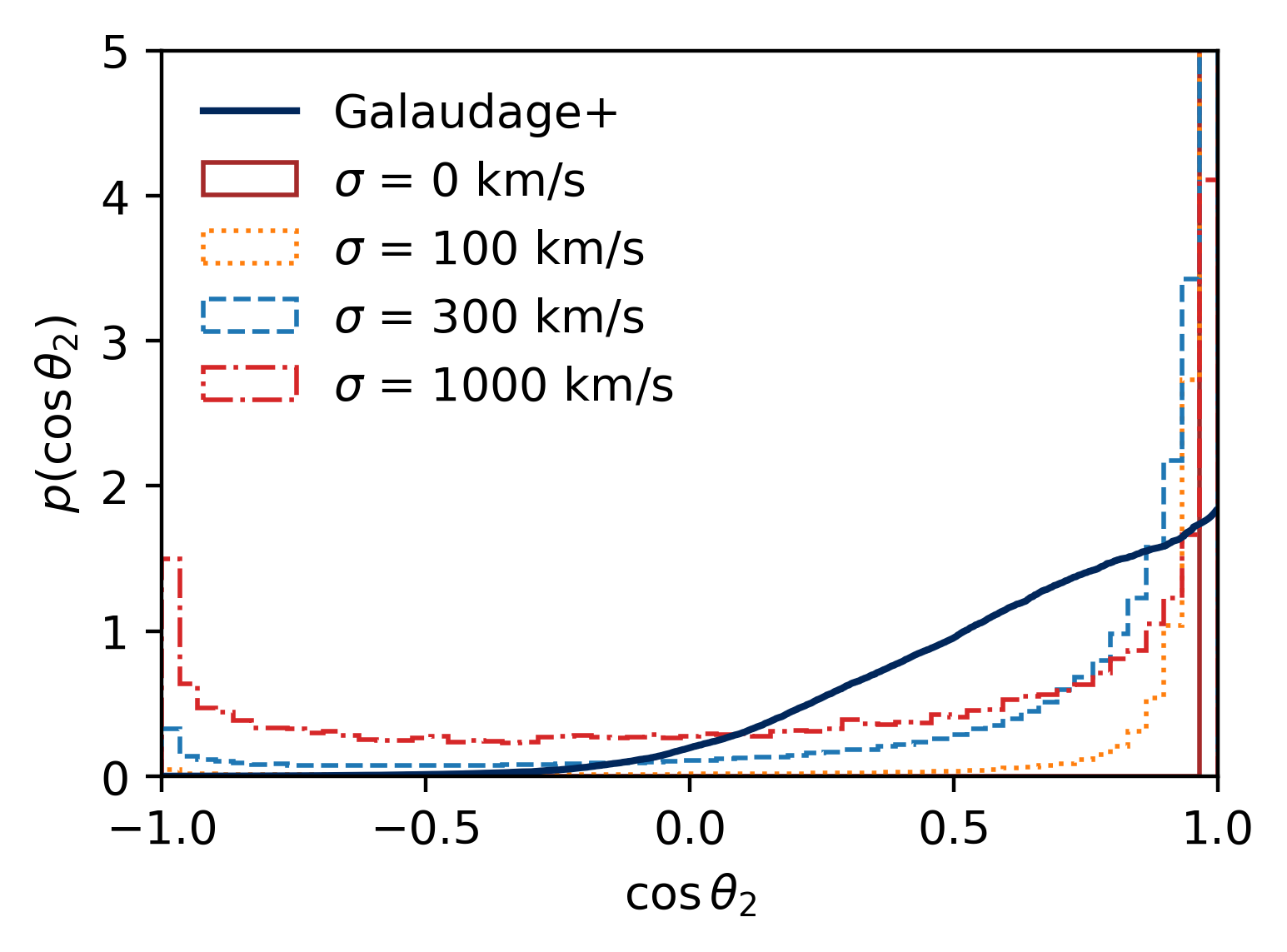}
    \caption{Normalised probability distribution for $\cos \theta_{2}$, the cosine of the angle between the spin of the secondary black hole and the binary orbital angular momentum, for several choices of $\sigma$. Also shown in the solid black curve is the fit to the gravitational-wave observations from \citet{Galaudage:2021rkt}.}
    \label{fig:costheta2}
\end{figure}

However, even though \citet{Galaudage:2021rkt} find no evidence for binary black holes with negative $\chi_\mathrm{eff}$, they still find some evidence for misaligned (precessing) black hole spins, with typical values for spin tilts of around $\cos \theta_{2} \sim 0.7$ (see Figure~\ref{fig:costheta2}).
If we assume that black hole kicks are the only mechanism capable of explaining this misalignment, then this implies that black holes receive kicks at birth.
We therefore allow for black hole kicks drawn from a Maxwellian with $\sigma = 100$, $300$ and $1000$\,km/s, motivated by observations of Galactic X-ray binaries (\citealp[e.g.,][]{Atri:2019fbx} and the results of \citet{Callister:2020vyz}; see also the discussion of other observational evidence for black hole kicks in \citealp{Callister:2020vyz}). 
We assume isotropically distributed kick directions, and compute the post-supernova orbital characteristics in the standard way \citep{Kalogera:1996ApJ,Callister:2020vyz}.
As stated above, we assume that the spins are aligned prior to the second supernova, such that the spins of both objects are misaligned by the same amount ($\cos \theta_{1} = \cos \theta_{2}$).
We show the distribution of misalignment angles $\cos \theta_{2}$ in Figure~\ref{fig:costheta2}. 
We find that kick distributions with $\sigma = 100$--$300$\,km/s produce qualitatively similar distributions of black hole spin-orbit misalignments to those inferred by \citet{Galaudage:2021rkt}, although the agreement is clearly not perfect. 
We have not tried adjusting other parameters in the \citet{Callister:2020vyz} model in order to improve the agreement; we suggest some avenues for future exploration in Section~\ref{sec:conclusion}.
Whilst most binaries still have their spins preferentially aligned with the orbital angular momentum ($\cos \theta_{2} \sim 1$), there is a tail of systems with misalignment angles $\cos \theta_{2} \lesssim 0.7.$
In this case, a small fraction of systems have $\cos\theta_{2} < 0$ and $\chi_\mathrm{eff} < 0$.
For larger values of $\sigma$ (for example, $\sigma = 1000$\,km/s) we find that a large fraction of binaries that remain bound have $\cos\theta_{2} < 0$, in contrast to the findings of \citet{Galaudage:2021rkt}.
While the tail of the $\cos \theta_{2}$ distribution is sensitive to the distribution of black hole kicks, we find that the $\chi_\mathrm{eff}$ distribution is not.

\section{Summary and conclusion}
\label{sec:conclusion}

A recent study by \citet{Callister:2020vyz} suggested that black holes must receive unexpectedly large kicks at birth (greater than 260\,km/s at 99\% confidence) to reconcile the observed spin distribution of merging binary black holes with the observed sample, assuming that all binaries were formed through the classic isolated binary evolution channel \citep{Belczynski:2016obo,Stevenson:2017tfq}. 
This is much higher than the typical kicks that black holes are expected to receive \citep[][]{Atri:2019fbx}, leading \citet{Callister:2020vyz} to consider alternate explanations.

We have argued that this may result arise due to an unfounded assumption made by \citet{Callister:2020vyz}. 
In particular, \citet{Callister:2020vyz} assume that the second formed black hole in a binary black hole system can be tidally spun up, regardless of the orbital separation.
Their best fit effective spin distribution has a Gaussian-like shape, peaked at slightly positive $\chi_\mathrm{eff}$, with a tail of negative $\chi_\mathrm{eff}$ systems, similar to the results obtained by \citet{LIGOScientific:2020kqk}.

In binaries with orbital periods greater than about 1\,day, tidal synchronisation is inefficient and slowly rotating black holes are formed \citep{Qin:2018vaa,Bavera:2019,Bavera:2021evk}.
We have modified the model from \citet{Callister:2020vyz} to account for this, using fits to detailed binary evolution models from \citet{Bavera:2021evk}.
We show that this leads to a bimodal spin magnitude distribution where most binary black holes will have $\chi_2 \sim 0$ (see Figure~\ref{fig:chi_2}).
It is unsurprising that the Gaussian spin model used by \citet{Callister:2020vyz} failed to capture such a distribution, and is likely the cause of their error.
The bimodal spin magnitude distribution also results in a more complicated effective spin distribution (see Figure~\ref{fig:chi_eff}) than found by \citet{Callister:2020vyz}. 
\citet{Galaudage:2021rkt} have recently shown that a similar effective spin distribution (as shown in Figure~\ref{fig:chi_eff}) provides a much better description (as quantified through the Bayesian evidence) of the observed binary black hole population than the model used by \citet{LIGOScientific:2020kqk}, and that there is no evidence for binaries with negative $\chi_\mathrm{eff}$ \citep[see also][]{Roulet:2021hcu}.
Thus the evidence for large black hole kicks claimed by \citet{Callister:2020vyz} may simply be the result of a model misspecification. 


We otherwise closely followed the assumptions made by \citet{Callister:2020vyz} regarding the properties of BH + He star binaries, and many of the caveats pertaining to that work also apply here.
For example, the distribution of orbital periods of BH + He star binaries can be determined by factors such as whether mass transfer is predominantly stable (typically predicting wider binaries) or unstable, leading to common envelope evolution \citep{Neijssel:2019,Bavera:2019,Bavera:2020uch,Olejak:2021A&A,Gallegos-Garcia:2021hti}.
As a demonstration of this we show the distribution of $\chi_
\mathrm{eff}$ for the stable mass transfer and common envelope channels from the population synthesis models of \citet{Bavera:2020uch} in the lower panel of Figure~\ref{fig:chi_eff}.


We used the fits to detailed binary evolution models from \citet{Bavera:2021evk} to determine the spin of the secondary black hole. 
However, these models have their own uncertainties, such as the assumptions made about the initial rotation rates (prior to tidal spin up) of the helium stars \citep{Qin:2018vaa,Bavera:2021evk,Renzo:2021}, and their mass-loss rates \citep{Detmers:2008A&A,Qin:2018vaa}.


We assumed that the distribution of black hole kicks is Maxwellian, following \citet{Callister:2020vyz}. 
However, this should be seen as a starting point, as there is no strong theoretical or observational motivation for a distribution of this form.
By the same token, a more complicated (e.g., bimodal) distribution is not justified by observations \citep[][]{Atri:2019fbx}. 
Hence, we opted to use a Maxwellian distribution primarily for reasons of simplicity and consistency. 
Future analyses may also wish to incorporate the dependence of black hole kicks on the properties of the collapsing star (such as the expected anti-correlation with mass; \citealp[e.g.,][]{Mandel:2020qwb}), although again, this may not be justified observationally \citep{Atri:2019fbx}.
If one were to include this theoretical expectation, one runs the risk of making the results strongly dependent on the model assumptions.

A logical next step to the analysis presented here would be to modify the phenomenological model from \citet{Callister:2020vyz} to account for the possibility of a bimodal spin magnitude distribution. There are several ways one might go about this. One could, for example, choose a similar model set up to that described by \citet{Callister:2020vyz}, but with two arbitrarily defined subpopulations with separately modelled spin magnitude distributions; those of `close' binaries in which tidal spin up is efficient, and `wide' binaries where it is not. 
Both our analysis and previous work suggest that an orbital period of 1\,day may be an appropriate dividing line between these subpopulations \citep{Qin:2018vaa,Bavera:2021evk}.
Alternatively, one could choose to use the physical relation between orbital separations and spin magnitudes (as encoded in the \citealp{Bavera:2021evk} fitting formulae) and instead fit the distribution of orbital periods. 
Regardless, one could then use such a model to infer the typical magnitude of black hole kicks with a similar method to 
\citet{Callister:2020vyz}.


In principle phenomenological models offer a method of examining the data with a model less dependent on specific choices made in modelling a particular astrophysical formation scenario.
However, we have shown here that they are not without issue.  
While population synthesis models naturally capture correlations between various quantities (such as a possible anti-correlation between black hole mass and kick mentioned above, \citealp[e.g.,][]{Mandel:2020qwb}), such correlations are usually neglected in phenomenological models such as the one discussed here. 
Another correlation that may be important to account for is that between the orbital separation and the masses in BH + He star binaries depending on whether they form through stable or unstable mass transfer \citep{Bavera:2020uch,Zevin:2020gbd,vanSon:2021}.
We have shown one example of how accounting for these correlations in a phenomenological model can influence the conclusions one draws from such a model.

In addition, this phenomenological model is broadly inspired by a single possible formation channel.
Population synthesis models predict that even under the umbrella of isolated binary evolution, there are multiple formation pathways for binary black holes \citep[e.g.,][]{Gerosa:2018wbw}, such as through double core common-envelope events \citep{Neijssel:2019,Olejak:2021iux}, through close binaries experiencing chemically homogeneous evolution \citep{Mandel:2015qlu,Marchant:2016wow} or through close binaries that experience mass transfer whilst the the primary is still on the main sequence, leading to rapidly rotating black holes \citep{Neijssel:2021imj}.
Each of these formation channels makes different predictions about the masses and orbital periods of BH + He star binaries.
Even different models of the same formation channel may not fully agree on predictions for the properties of BH + He star binaries due to the many uncertainties in population synthesis models \citep[e.g.,][]{Gallegos-Garcia:2021hti}.
Future phenomenological models will need to account for these pathways in order to obtain unbiased inferences. 

There is currently no evidence for black holes receiving large natal kicks ($\sim 1000$\,km/s) in the spin distribution of merging binary black holes observed through gravitational-waves. 
The population is consistent with most binary black holes having their spins (approximately) aligned with the binary orbital angular momentum (cf. Figure~\ref{fig:costheta2}).
Given that some exceptional events such as GW190521 \citep{Abbott:2020tfl,Abbott:2020mjq} and GW190814 \citep{LIGOScientific:2020zkf} have properties that are difficult to explain through isolated binary evolution, there may be contributions to the binary black hole population from other formation channels \citep{Bavera:2020uch,Zevin:2020gbd}. 
For the remainder of the population, we have shown that at least the overall spin distribution is explainable through isolated binary evolution.
There is preliminary evidence that black holes receive kicks of order $\sim 100$\,km/s.
In the future, it may be possible to use properties of the merging binary black hole population, such as the overall merger rate, along with the mass and spin distributions to constrain the magnitude of black hole kicks \citep[e.g.,][]{Stevenson:2015bqa,Zevin:2017evb,Barrett:2017fcw,Wong:2019uni}.


\acknowledgments

We thank Tom Callister and Mathieu Renzo for engaging discussions on this topic, and Ilya Mandel, Shanika Galaudage and Eric Thrane for useful comments on the manuscript.
We additionally thank the referee for constructive criticism that helped improve the paper.
SS is supported by the Australian Research Council Centre of Excellence for Gravitational Wave Discovery (OzGrav), through project number CE170100004. 
SS is supported by the Australian Research Council Discovery Early Career Research Award through project number DE220100241.
\newline


\section*{Data availability}

The results from \citet{Callister:2020vyz} are publicly available at \href {https://github.com/tcallister/state-of-the-field-gwtc2}{this url}.
The results from \citet{Galaudage:2021rkt} are publicly available at \href{https://github.com/shanikagalaudage/bbh_spin}{this url}.
The results from \citet{Bavera:2021evk} are publicly available at \href{https://github.com/ssbvr/approximating_BH_spins}{this url}.
The results from \citet{Bavera:2020uch}, as released by \citet{Zevin:2020gbd}, are publicly available at \href{https://zenodo.org/record/4947741#.YTwCFMauanI}{this url}.
All data produced in this work is available from the author upon request.

%




\bibliographystyle{mnras}
\bibliography{bib} 

\end{document}